# Fractal Optimization as Generator of Market Neutral Long-Short Portfolio

Sergey Kamenshchikov, PhD, *IFCM Group/ Moscow Exchange*
Ilia Drozdov, CFA, *QB Capital*

**Abstract**

*A fractal approach to the long-short portfolio optimization is proposed. The algorithmic system based on the composition of market-neutral spreads into a single entity was considered. The core of the optimization scheme is a fractal walk model of returns, optimizing a risk aversion according to the investment horizon. The covariance matrix of spread returns has been used for the optimization and modified according to the Hurst stability analysis. Out-of-sample performance data has been represented for the space of exchange traded funds in five period time period of observation. The considered portfolio system has turned out to be statistically more stable than a passive investment into benchmark with higher risk adjusted cumulative return over the observed period.*

**Introduction**

According to the research of Malkiel [1] only 14% of long-term equity funds represent an average return of 2-4% above S&P500 benchmark in ten years time frame. This statistics correspond to pre-ETF era of 1990-2001. However the typical Sharpe ratio of S&P500 reaches 1.5-2 levels only for the 5-10 years horizon. In combination with bond funds it makes S&P500 index a comfortable instrument for pension programs, but inefficient for middle-term investment of 1-5 years horizon. The local diversification of Long-Only funds doesn't efficiently provide a systematic risk aversion. The Global Asset Allocation models simplified a diversification at the beginning. However while there are several major drivers of the global Market like the US or Asia this model still lacks a market-neutrality in long term strategies. Another approach to market neutral investment is a portfolio of hedge funds which apply short term long-short arbitrage models with high beta neutrality. Unfortunately hedge fund models still preserve properties of "black boxes" and are not comfortable for the transparent investing. In the current research we prove that a long-short model is suitable for a long term investment and may provide stable trends. This model is based on market-neutral pair spreads which use relative competitive advantages of assets. Diversification of spreads allows eliminating both systematic and non systematic portfolio risks. We introduce a fractal model of volatility to account for nonlinear risks such as volatility clustering. This approach suggests a new step outside the standard statistics. In following sections we provide description of market neutralization of spreads, their composition into the single entity and portfolio optimization.

**Market neutrality**

Let's represent daily returns of assets *i* and *j* in the linear form:

$$r_i(t) = r_i^0 + \beta_i r_m(t) + \varepsilon_i(r_m) \qquad r_j(t) = r_j^0 + \beta_j r_m(t) + \varepsilon_j(r_m) \qquad (1)$$

Here $r_i^0$ and $r_j^0$ are constant drift terms, $r_m$ is a mutual market return. Residuals express a random component in case of a perfect regression model. Otherwise residuals may be represented as nonlinear functions of market returns. If we consider long-term investments, returns are to be normalized in relation to the investment entry point at the beginning of the holding period:

$$r_i(t) = \frac{p_i(t) - p_i(t-_\Delta t)}{p_i^0} \qquad r_j(t) = \frac{p_j(t) - p_j(t-_\Delta t)}{p_j^0} \qquad (2)$$

The market term expresses the mutual market of these two assets which has to be defined in a quantitative way. Consequently $\beta_i$ and $\beta_j$ are constant factors that show a relation of each asset to the market linear motion.

Definition of betas may be expressed through the increments of returns:

$$\langle \delta r_i(t) \rangle = \beta_i \langle \delta r_m(t) \rangle + \xi_i(r_m) \quad \langle \delta r_j(t) \rangle = \beta_j \langle \delta r_m(t) \rangle + \xi_j(r_m) \tag{3}$$

Here new residuals are weakly nonlinear terms $\xi_{i,j}(r_m) = \langle \delta \varepsilon_{i,j}(r_m) \rangle$. Hedge factors $\chi_{ij} = \beta_i / \beta_j$ may be defined by the relation with weakly nonlinear residual (4):

$$\overline{\chi}_{ij} = \frac{\langle \delta r_i(t) \rangle}{\langle \delta r_j(t) \rangle} = \chi_{ij} + \eta_{ij}(r_m, \chi_{ij}) = \frac{\beta_i + \langle \delta \varepsilon_i \rangle / \langle \delta r_m \rangle}{\beta_j + \langle \delta \varepsilon_j \rangle / \langle \delta r_m \rangle} \tag{4}$$

Here $\eta(r_m, \chi_{ij})$ is a weakly nonlinear term. Long/Short position (spread) of assets $i$ and $j$ correspondingly tends to the perfect market neutral state while two conditions are satisfied: $w_i / w_j = 1 / \chi_{ij}, \eta \to 0$. Here $w_i, w_j$ are relative weights of assets $i$ and $j$ correspondingly. Second condition is equivalent to $\langle \delta \varepsilon_{ij} \rangle \prec\prec \langle \delta r_m \rangle$. The return of this spread may be represented as the superposition of constant term and weakly nonlinear term $\Omega$:

$$\Delta_{ij} = r_i(t) - \overline{\chi}_{ij} r_j(t) = \left( r_i^0 - \chi_{ij} r_j^0 \right) + \Omega_{ij}(r_m, \chi_{ij}) = const + \Omega_{ij}(r_m, \chi_{ij}) \tag{5}$$

A fundamental sense of a stable spread return is a competitive advantage of the asset $i$ in comparison to the asset $j$ with weight correction. One way to reach perfect market neutrality is to make regression tests, such as Fisher test, based on the standard statistics. However, a standard statistics and normal distribution of residuals doesn't take into account nonlinear effects such as clustering of volatility, "three-sigma" events, which provide additional risks [2]. In the current research we propose a fractal walk model of returns for analysis of volatility $\Omega$ - effects in the stage of spread selection and portfolio optimization.

**Spread selection and ranging**

The basic idea of proposed portfolio is a composition of beta-neutral spreads into single entity that can be transformed into long/short investment portfolio. However there is uncertainty in the stage of a spread selection, which can be eliminated by the ranging mechanism. For $N$ assets we work with upper triangular part of a generating matrix $\{\Delta_{ij}\}, i = \overline{1, N}; j = \overline{1, N}$. Assets, activated in a spread $(i, j)$ are removed from the generating space to provide diversification of spread returns. Therefore a correct consequence of spread activation is an important part of the whole portfolio composition. Below we propose a criterion of evolutionary spread selection, based on the fractal walk hypothesis. A standard model of a geometric random walk, Gaussian process, implies normally distributed volatility of spread returns:

$$\Delta_{ij}(t) = \Delta_{ij}^0 + \theta_{ij} \, _\Delta W(t) \tag{6}$$

Amplitude of volatility $\theta_{ij}$ is assumed to be a constant factor. Volatility function is represented by the normally distributed Wiener random walk $_\Delta W(t) \infty N(0,1)$. According to the Kelly representation [3] an optimal relative weight $w_{ij}$ may expressed from the return growth maximum condition. Averaging is realised according to the horizon $_\Delta t$:

$$\left\langle \frac{_\Delta \Delta_{ij}}{_\Delta t} \right\rangle = w_{ij} \Delta_{ij}^0 - \frac{w_{ij}^2 \theta_{ij}^2}{2} \to \max \tag{7}$$

This criterion defines an optimal relative weight of the portfolio spread if cross correlation effects are neglected: $w_{ij} = \mu_{ij} / (\theta_{ij})^2$. The term $\theta_{ij}$ is estimated as a variance of returns. This estimation is sufficient for a one-period (one day) investment horizon.

For *N* days a rescaling of Gaussian process gives the following estimation:

$$w_{ij} = \frac{\Delta_{ij}^0 N}{\left(\theta_{ij}\right)^2 N^2} = \frac{\Delta_{ij}^0}{\left(\theta_{ij}\right)^2 N} \quad (8)$$

However a constant term *N* may be neglected for the task of relative weighting – the investment horizon is excluded from the parametric space. In this point a standard random walk approach is to be modified. In this research we propose a one-parameter model of volatility evolution – a model of Fractal Brownian Motion (FBM), introduced by Kolmogorov and Mandelbrot [4]. This model is relatively simple (one parameter) but explains such effects as volatility clustering and instabilities [5]. It may be applied as a second estimation model without significant processing resources. According to Mandelbrot [4] the FBM of the exponent $H$ is a moving average of $_\Delta W(t)$ with increments weighted by the kernel $(_\Delta t)^{H-1/2}$. Let's denote a basic relation for the volatility rescaling, derived from the FBM model:

$$E\left(\left[\Delta_{ij}(t_2) - \Delta_{ij}(t_1)\right]^2\right) = |t_2 - t_1|^{2H} \quad (9)$$

Expectation for the price fluctuation square in two arbitrary time moments $t_1, t_2 : t_2 \succ t_1$ is rescaled through the Hurst exponent, which is a measure of time series memory. Correspondingly an average one period volatility $\theta_{ij}^1$ may be used to express the *N*-periods volatility $\theta_{ij}^N$ in the following way: $\theta_{ij}^N = \theta_{ij}^1 \cdot N^{H_{ij}}$. For the standard random walk process a Hurst exponent has a fixed value of 0.5 and doesn't take into account evolution of volatility due to non stationary effects. The proposed model allows modifying of optimal weight with account for an individual Hurst exponent of spread $H_{ij}$:

$$w_{ij} = \frac{\Delta_{ij}^0 N}{\left(\theta_{ij}\right)^2 N^{2H_{ij}}} = \frac{\Delta_{ij}^0}{\left(\theta_{ij}\right)^2 N^{1-2H_{ij}}} \quad (10)$$

An estimation of $\Delta_{ij}^0$ is an average daily return of spreads $i, j$. A definition of Hurst exponent is an object of numerous researches. In this paper we apply an approach of small data basis, suggested in [6]. This algorithm gives an estimated Hurst exponent $H_{ij}$ and its error $_\Delta H_{ij}$. We should recall that the mean-reversion process corresponds to area of parameters $H_{ij} \in (0, 0.5)$. The critical level $H_{ij} = 0.5$ may be decreased for the searching of more stable spreads. Yet, this procedure leads to narrowing of spread space and decrease in diversification. This critical level may be typically used for the empirical optimization of current portfolio. In contrast we propose only a prototype optimization with maximal critical level $H_{ij} = 0.5$.

A preliminary selection of spreads may be executed on the basis of a following algorithm:
- The calculation of maximal weight in space of spreads $\{\Delta_{ij}\}$
- The control for the criterion $H_{ij} + {_\Delta H_{ij}} \prec 0.5$ and $_\Delta H_{ij} \prec H_{ij}$
- A record of corresponding spread assets $i, j$ and calculation of portfolio spread $\Delta_{ij}$ or $\Delta_{ij}$ according to the condition $\langle \Delta \rangle \succ 0$
- Exclusion of assets $i, j$ from the generating spreads
- Return to step one

As a result, we prefer spreads with $\Delta_{ij}(t) \to \Delta_{ij}^0$ over more volatile components. The final space of spreads can be used for more complex nonlinear portfolio optimization. In this stage we prepared a set of "long" positions for spreads $k = \overline{1, M}$.

**Portfolio optimization**

According to the Kelly interpretation [3] effects of cross correlation may be taken into account by covariance matrix of returns: $C = \text{cov}(\Delta_l, \Delta_r)$, $l = \overline{1, M}$, $r = \overline{1, M}$. Here we consider spreads as the synthetic market-neutral assets with "long" position. A standard random walk model allows defining relative weights vector in accordance with the standard solution: $\vec{w} = C^{-1} \vec{\Delta}$. Here we consider modification according the fractal walk approach, described above. An element of covariance matrix $\text{cov}(\Delta_l, \Delta_r)$ can be modified according to the definition of correlation:

$$\text{cov}(\Delta_l, \Delta_r) = \text{cov}(_\Delta W_F^l, _\Delta W_F^r) = \theta_l \theta_r R_{lr} \qquad (11)$$

Here $_\Delta W_F^l$ are fractal motion residuals of the modified model (6) while $R_{lr}$ is a correlation factor. Relations for daily volatilities can be rescaled for the $N$ horizon: $\theta_l \to \theta_l \cdot N^{H_l}$, $\theta_r \to \theta_r \cdot N^{H_r}$. A basic question is rescaling of correlation – a squared root of determination factor for linear regression model $\Delta_l = k_{lr} \Delta_r + \xi_l$. The definition gives us a statistical relation of $R_{lr}$:

$$R_{lr} = \left( \frac{Var(\Delta_l^*)}{Var(\Delta_l)} \right)^{1/2} \to R_{lr} \cdot N^{1-H_l} \qquad (12)$$

Here variance of linear term $\Delta_l^* = k_{lr} \Delta_r$ is compared to the total variance $Var(\Delta_l) = \theta_l^2$. Rescaling of both variances allows finding a rescaling law for the correlation. A linear regression curve has a Hurst factor $H_l^* = 1$. Therefore a final solution can be represented as a right part of (12). Here a non rescaled correlation is to be derived from the input data. A final relation for the rescaled covariance element is represented in the following way (arrow symbol designates rescaling):

$$C_R = \{\text{cov}(\Delta_l, \Delta_r)\} \to \{\text{cov}(\Delta_l, \Delta_r)\} \cdot N^{H_r+1} \qquad (13)$$

The criterion of allocation is a simple consequence of this relation:

$$\vec{w}_R = C_R^{-1} \cdot \vec{\Delta} \cdot N \qquad (14)$$

In a portfolio investment $N$ corresponds to the investment horizon – a portfolio manager parameter. In this research we consider a period of half year. An algorithm of optimization may be represented in the following way:

- Definition of daily volatilities of spreads $\theta_{l,r}$
- Definition of non rescaled covariance matrix $C = \text{cov}(\Delta_l, \Delta_r)$
- Rescaling of covariance matrix and volatilities
- Calculation of rescaled covariance matrix $C_r$
- Calculation of rescaled vector of relative weights $\vec{w}_R$
- Composition of long/short positions
- For leverage $l$ the final vector $\vec{w}_R$ is multiplied by constant factor: $k = l / \sum_n w_R^n$

Leverage l is a measure of risk tolerance – from conservative to aggressive portfolio. We consider l=2 in this research.

**Input data**

In current investigation we use the universe of most liquid equity-based exchange traded funds, ETFs. We take into account the growing popularity of ETFs among investment firms due to their internal diversification and relatively low costs unlike these of mutual funds. According to [7] an average growth of ETF AuM (assets under management) is 27% per year for recent 10 years.

High liquidity of chosen ETFs gives opportunity to make realistic simulation of the current portfolio system and to decrease impact of trade fees. The liquidity has been estimated through the average half-year volume: May 2016-October 2016. The algorithm of ETFs filtering is given below:

- 100 most liquid ETFs are chosen as the initial space
- ETFs introduced later than 2011 are excluded from the initial space
- ETFs other than passively managed are excluded
- Non-leveraged ETFs are chosen
- Reverse ETFs are excluded
- Volatility ETFs are excluded
- Non-shortable ETFs are excluded

We prefer passively managed ETFs in order to minimize subjective properties of management team – the current research is sufficiently quantitative. Any artificial influences like risk control, leveraging are excluded. We tend to use assets that reflect purely market trends – reverse and volatility ETFs are excluded for this reason. The final list includes 25 funds: EEM, XLF, IWM, QQQ, EFA, XLU, XLP, XLI, XLV, XLK, AMLP, ITB, XBI, KRE, XLY, XLB, EWW, XME, EZY, EWT, KBE, EWH, EWY, EWY, XHB.

We limit our out-of-sample simulation to five years – a subjective consensus between diversification of the universe and the historical significance. Adjusted daily close prices are used - exported from Interactive Brokers (IB) historical datebase. Commissions and overnight rates correspond to IB trading conditions as well: commission of $0,005 for single stock, overnight rate of 1% per annum. Selected ETFs are quoted in US dollars and traded in NYSE/NASDAQ. Internal management fees are added according to individual conditions of ETF issuers. The considered initial investment is $100,000 which corresponds to the institutional level of capital - however we admit the utilization of this model to the capital above $50,000.

**Simulation results**

Simulation is based on historical adjusted closing prices for November 2011-Novermber 2016. We use out-of-sample testing for the model validation. The frame of training data corresponds to half year – the same as out-of-sample horizon. The consequence of testing may be represented in the following way:

- Retrieving prices for the first half year training sample
- Calculation of investment weights vector and portfolio composition
- Testing portfolio performance for the next half year test sample
- Recording performance metrics
- Using current test sample as the training data

The overall observation period includes five years frame. We compare the performance with buy-and-hold passive strategy with SPY being a benchmark. In Fig.1a the cumulative return of reinvestment strategy is represented for ten periods (five years, half year horizon) with leverage $l=2$. The correlation of half-year returns and SPY returns corresponds to 24% level, which confirms the efficient hedging of proposed strategy. Diversification is preserved: the number of assets varies from 14 to 20 with average investment weight of 19%. The average return in frame of reinvestment scheme is higher for benchmark (market) – 11% against 10% for portfolio.

However the $R^2$ of portfolio linear regression is 92%, 5% higher than one for passive SPY. It may be useful to analyze stability in relation to risk measure, volatility. We use single half-year returns without reinvestment for estimation of an average single return: 4.3% for portfolio and 4.6% for benchmark. Annual single returns estimations are 8.6% and 9.2% correspondingly.

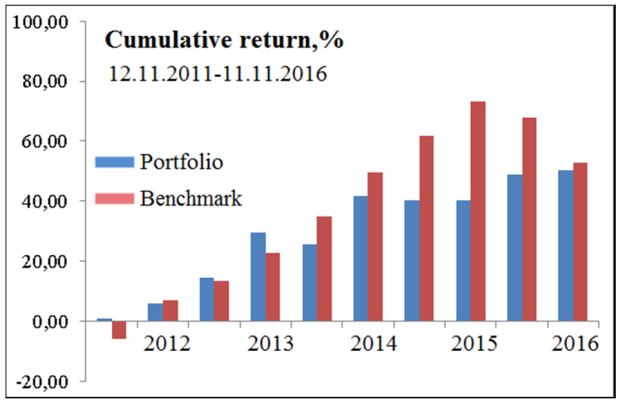 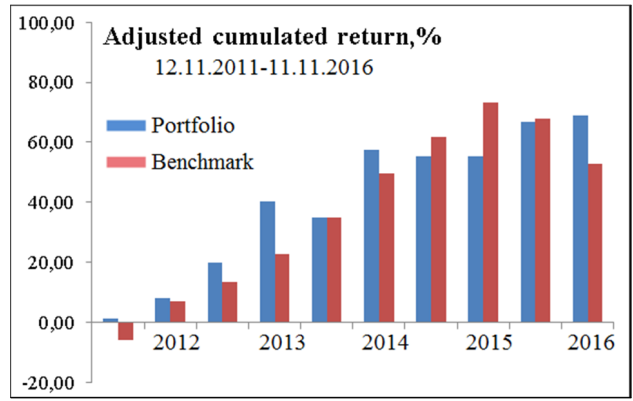

**Fig. 1a. Cumulative returns, half year horizon**   **Fig. 1b. Risk adjusted cumulative returns**

Half year volatilities (ten points) may be approximately rescaled into annual volatilities according to the Lo relation $V \to V\sqrt{2}$ [8]. Finally we derived Sharpe of 1.07 for portfolio system and 0.84 for benchmark investment. The comparative performance metrics are represented in Table 1.

|  | Portfolio | Benchmark |
|---|---|---|
| **Volatility,%** | 133 | 170 |
| **Max. drawdown,%** | 12 | 21 |
| **Market neutrality,%** | 76 | 0 |
| **Annual return (reinvested),%** | 10 | 11 |
| **Annual return,%** | 8.6 | 9.2 |
| **Annual Sharpe ratio** | 1.07 | 0.84 |

**Table 1. Comparative performance metrics**

Volatility is calculated among individual half-year returns and is normalized in relation to average half-year return. We recorded a maximal weight in each out-of-sample test and calculated an average quantity, represented in the first raw of the table. Maximal drawdown is calculated on the basis of cumulative daily returns (Fig.2). Here we apply the relation of a half-year volatility of market and the portfolio strategy. The result is represented in Fig.1b. This result corresponds approximately to the leverage *l*=2.7 and is too risky as it provides drawdown of 17% of initial capital (Fig.2, rectangular).

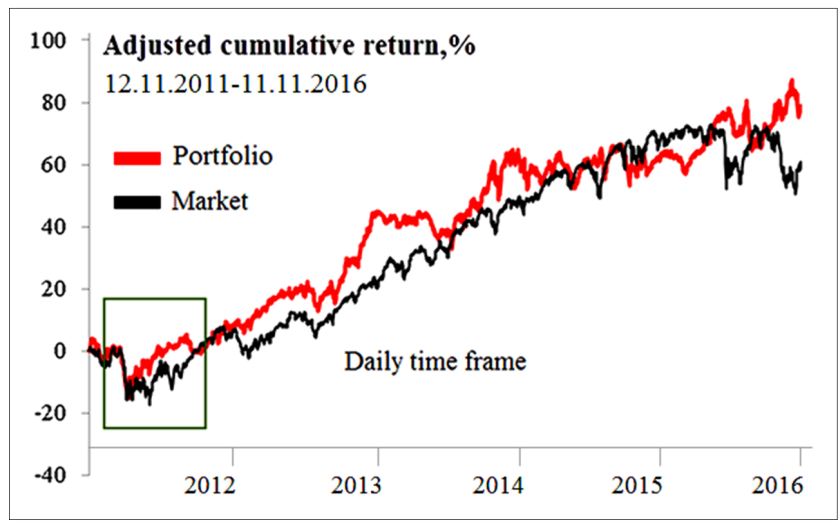

**Fig. 2. Cumulative returns, daily horizon, *l*=2.7**

The adjusted annual cumulative returns may be calculated on the basis of half-year frames (Fig. 3a). The portfolio system outperforms a passive investment for all periods except 2015. In 2014 the difference of cumulative return decreased as well by 10%. While single returns have sufficiently decreased in 2015 (Fig. 3b), we observe the reversion to the average return in 2016.

We suppose that a restructuring of the oil market in 2014-2015 could influence the competitive relations between spread ETFs. However the optimization of portfolio at the end of the year took into account these new statistical properties. We suppose that careful monitoring of portfolio rebalancing may decrease this effect.

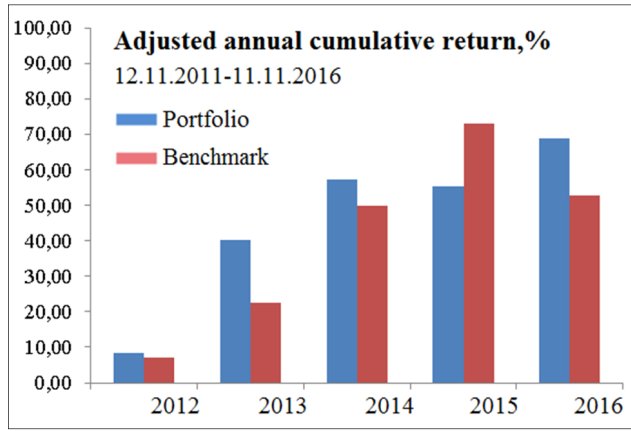 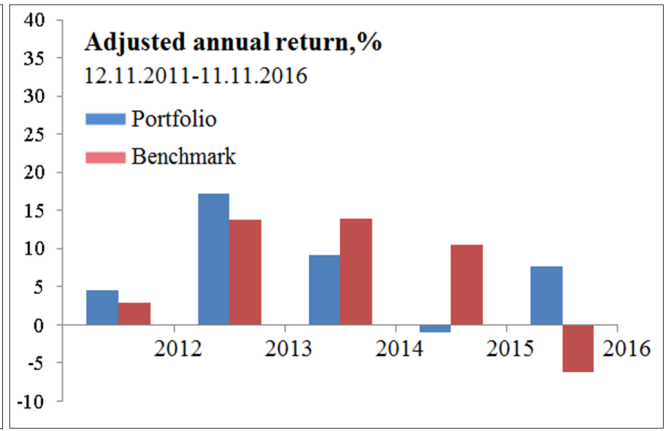

Fig. 3a. Annual cumulative returns, *l*=2.7    Fig. 3b. Annual single returns, *l*=2.7

The following remarks can be made on the basis of the represented data:

- Portfolio system is stable enough for one and two years horizon
- Its risk adjusted cumulative return is 1.3 times higher than benchmark return
- Its neutrality is significant enough to use it as a hedging opportunity

**Conclusions**

In this paper we applied a hybrid approach to the portfolio risk elimination. Portfolio algorithmic system is based on composition of market-neutral pair spreads of ETFs into single diversified entity. Hedging of market risks is achieved in frame of a linear regression model for return increments. Control of model validity is realized through the fractal walk model of returns, which allows modifying risk analysis according to the constant investment horizon and possible volatility clustering. This model has been used for the preliminary selection of spreads from the generating matrix. The final space of spreads has been used for fractal portfolio optimization. It has been shown that the covariance matrix of returns should be rescaled according to the preferred investment horizon. Individual properties of spread returns have been taken into account through the Hurst stability analysis. Out-of-sample analysis has been realized for the space of 25 equity exchange traded funds (ETFs) during five-year period. We have compared the portfolio performance with a buy-and-hold passive strategy for SPY benchmark.

Hedging opportunities have been confirmed: correlation of half-year returns and SPY returns corresponds to 24%. Global diversification has been preserved as well: the number of assets varies from 14 to 20 with average investment weight of 19%. It turned out that a portfolio system is more stable than a passive investment for one year horizon. Its risk adjusted cumulative return is 1.3 times higher than benchmark return, which gives opportunities to beat the market with comparable risk tolerance. We observed a drawdown of return in 2015, which can be corrected by the additional rebalancing schemes. The advantage of this model is its internal sensitivity to investment horizon, which makes it comfortable for individual approach of portfolio advisors and gives protection from nonlinear risks. However we admit that the growth of passive investment may lead to the decrease of competitive difference between ETFs due to their growing beta. Besides, we understand the risks of regulations regarding short positions, particularly during financial crisis or recession. Nonetheless, the model may serve as a valuable diversifying asset that improves overall portfolio performance. It may be used as a prototype for more complicated models, including standards statistical tests.

# References

1. Malkiel, "Passive Investment Strategies and Efficient Markets", European Financial Management,Vol.9, No.1, 2003
2. Engle, Patton, "What good is a volatility model?", Quantitative Finance, Vol.1, 2001
3. Laureti, Medo, Zhang, "Analysis of Kelly-optimal portfolios", Quantitative Finance, Vol.7, 2010
4. Mandelbrot, Van Ness, "Fractional Brownian motions, fractional noises and applications", SIAM Review, Vol.10, 1968
5. Calve, Adlai, "The Review of Economics and Statistics", Vol.34, 3, 2002
6. Dubovikov, Starchenko, "Dimension of the minimal cover and fractal analysis of time series", Physica A, Vol.339, 2004
7. Bolla, Kohler, Wittig, "Index-Linked Investing—A Curse for the Stability of Financial Markets around the Globe?", The Journal of Portfolio Management, Vol. 42, No. 3, 2016
8. Lo, "The Statistics of Sharpe Ratios", Financial Analysts Journal, Vol.58(4), 2003

*Note: the paper is accepted in Quantitative Finance*